\documentclass[aip, apl, superscriptaddress, preprint]{revtex4-1}

\usepackage{graphicx}

\usepackage{color}
\usepackage{epstopdf}

\begin{document}

\title{Direct evidence of ferromagnetism in a quantum anomalous Hall system}

\date{\today}

\author{Wenbo Wang}
\affiliation{Department of Physics and Astronomy, Rutgers University, Piscataway, New Jersey 08854, USA}
\author{Yunbo Ou}
\affiliation{State Key Laboratory of Low Dimensional Quantum Physics, Department of Physics, Tsinghua University, Beijing 100084, China}
\author{Chang Liu}
\affiliation{State Key Laboratory of Low Dimensional Quantum Physics, Department of Physics, Tsinghua University, Beijing 100084, China}
\author{Yayu Wang}
\affiliation{State Key Laboratory of Low Dimensional Quantum Physics, Department of Physics, Tsinghua University, Beijing 100084, China}
\affiliation{Collaborative Innovation Center of Quantum Matter, Beijing 100084, P. R. China}
\author{Ke He}
\affiliation{State Key Laboratory of Low Dimensional Quantum Physics, Department of Physics, Tsinghua University, Beijing 100084, China}
\affiliation{Collaborative Innovation Center of Quantum Matter, Beijing 100084, P. R. China}
\author{Qi-kun Xue}
\affiliation{State Key Laboratory of Low Dimensional Quantum Physics, Department of Physics, Tsinghua University, Beijing 100084, China}
\affiliation{Collaborative Innovation Center of Quantum Matter, Beijing 100084, P. R. China}
\author{Weida Wu}
\email[Corresponding author: ]{wdwu@physics.rutgers.edu}
\affiliation{Department of Physics and Astronomy, Rutgers University, Piscataway, New Jersey 08854, USA}

\begin{abstract}

The quantum anomalous Hall (QAH) systems are of great fundamental interest and of potential application because of dissipationless conduction without external magnetic field\cite{haldane88, onoda03, liu08, qi08, yu10, qiao10, nomura11, zhang12, ezawa12}. The QAH effect has been realized in  
magnetically doped topological insulator thin films\cite{chang13,checkelsky14,Kou14prl,kou15,feng15}. However, full quantization requires extremely low temperature ($T< 50$\,mK) in the initial works, though it was significantly improved with modulation doping or co-doping of magnetic elements\cite{mogi15, ou17}. Improved ferromagnetism was indicated in these thin films, yet a direct evidence of long-range ferromagnetic order is lacking.  Herein, we present direct visualization of long-range ferromagnetic order in thin films of Cr and V co-doped (Bi,Sb)$_2$Te$_3$ using low-temperature magnetic force microscopy with \textit{in-situ} transport. The magnetization reversal process reveals a typical ferromagnetic domain behavior, \textit{i.e.}, domain nucleation and domain wall propagation,  in contrast to much weaker magnetic signals observed in the end members, possibly due to superparamagnetic behavior\cite{Lachmane15,grauer15,lee15}. The gate dependence of magnetic reversal indicates a significant role of bulk carrier-mediated exchange interactions. The observed long-range ferromagnetic order resolves one of the major challenges in QAH systems, and paves the way to high-temperature dissipationless conduction by exploring magnetic topological insulators.  

\end{abstract}

\maketitle
Dissipationless conduction is technologically appealing because of a wide range of potential applications. There are two ways to achieve dissipationless conduction in condensed matter systems, superconductivity and topological chiral edge states.  Therefore, high-temperature superconductivity, a great example of dissipationless conduction, has been extensively investigated for decades\cite{bednorz86,wu87,maeda88, schilling93}. Dissipationless conduction due to chiral edge states is realized in quantum Hall effect (QHE), which requires low temperature and external magnetic field. The closely related phenomenon, quantum anomalous Hall effect (QAHE) doesn't need an external magnetic field, so it attracts significant attention recently. The realization of QAHE requires time reversal symmetry breaking and topologically nontrivial band structure. There have been a few theoretical proposals\,\cite{haldane88, onoda03, liu08, qi08, yu10, qiao10, nomura11, zhang12, ezawa12}, and eventually it is experimentally realized in magnetically doped 3D topological insulator (TI) thin films\,\cite{chang13,checkelsky14,Kou14prl,chang15}. Here the ferromagnetism was introduced by doping magnetic elements. The broken time reversal symmetry opens a mass gap at the Dirac point of the topological surface states. By tuning the Fermi level inside the mass gap, the magnetic TI thin film is equivalent to two copies of half integer QHE systems\cite{Grauer2017}.  QAHE was first observed in Cr-doped Bi$_x$Sb$_{2-x}$Te$_3$ (BST) thin films synthesized by molecular beam epitaxy (MBE)\,\cite{chang13}. The quantized Hall conduction, however, was observed at ultra-low temperature ($\sim 30$\,mK). This observation is soon confirmed by other groups\,\cite{checkelsky14,Kou14prl,kou15,feng15}. Later, a robust QAHE with higher precision quantization was observed in the V-doped BST thin film, which is a hard ferromagnet with a larger coercive field ($H_c$) and higher Curie temperature ($T_\mathrm{C}$) with the same doping level\,\cite{chang15}. Yet, ultra-low temperature ($T<50$\,mK) is still needed to achieve full quantization. Therefore, it is imperative to understand the origin of the ultra-low temperature of full quantization, which is still under debate. 

Magnetic inhomogeneity has been proposed to be one of the main factors that limit the QAH temperature. The electronic inhomogeneity, disordered ferromagnetic or superparamagnetic behavior, have been reported in Cr-doped BST thin films\cite{chang14, lee15, Lachmane15,grauer15}. In these cases, the reduced QAH temperature is likely limited by the regions with the smallest exchange gap in the Cr-doped systems. While modulation doping of Cr was shown to improve the quantization temperature in penta-layer thin films\,\cite{mogi15}, it is unclear whether it reduces magnetic inhomogeneity.  On the other hand, recent angle-resolved photoemission spectroscopy (ARPES) studies of V-doped BST thin films suggest that valance band maximum (VBM) is above the Dirac point\cite{li16} so that ultra-low temperature and disorders are needed to localize the bulk carriers.  The different mechanisms of reduced quantization temperature indicate that  Cr and V co-doping might help to enhance the performance. Furthermore, alloy doping is commonly known as an effective route to improve ferromagnetic order in a diluted magnetic semiconductor\cite{andriotis13, qi16}.  

\begin{figure*}[htbp]
\includegraphics[width=0.8\columnwidth]{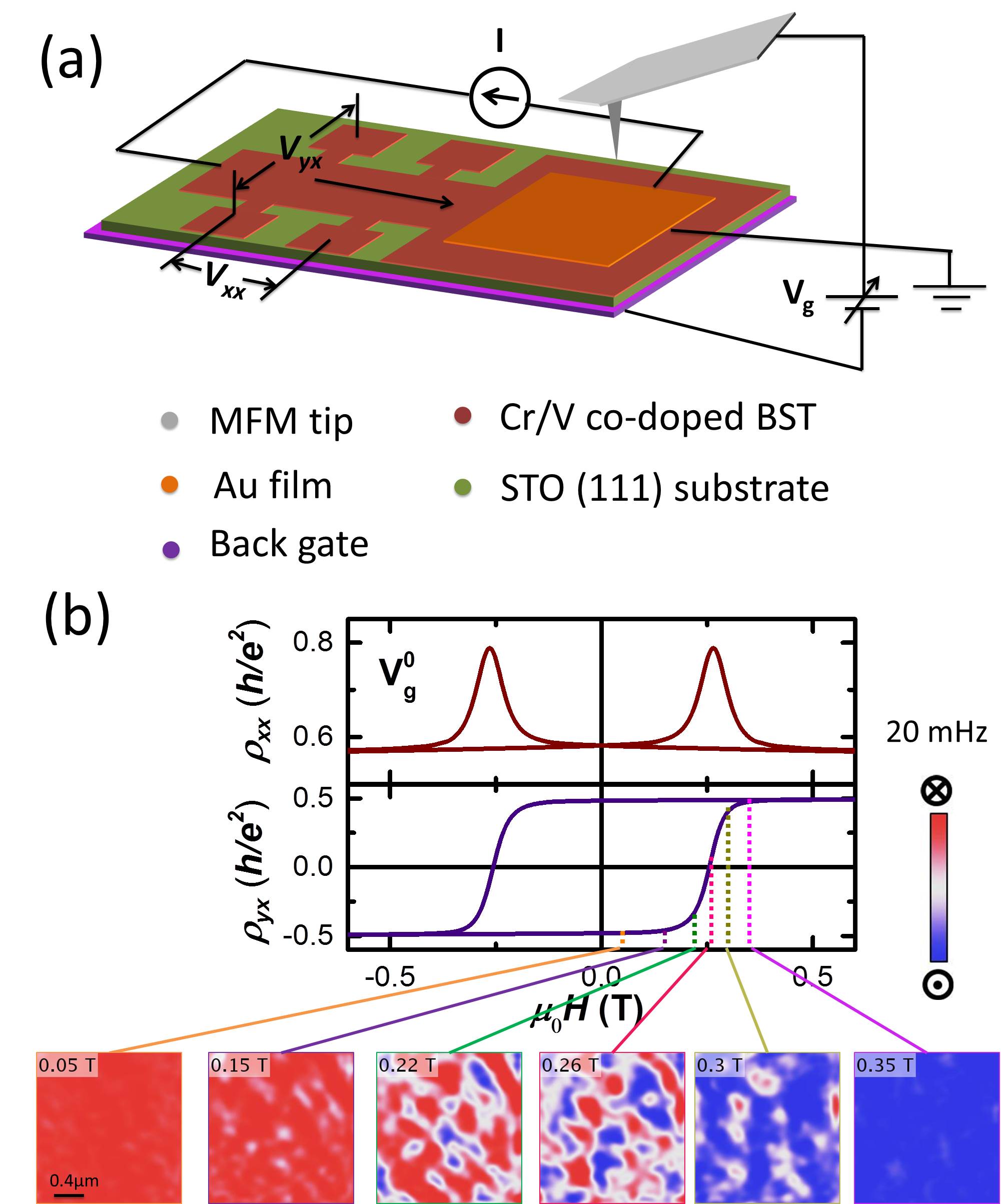}
\caption{\textbf{Schematic of  \textit{in-situ} transport setup and MFM data of the Cr/V co-doped BST film at 5\,K} $|$ (a) A cartoon of the Hall bar device for MFM and $\textit{in-situ}$ transport measurements. The 5\,QL Cr/V co-doped BST thin film was grown on STO(111) substrate using MBE, followed by deposition of a layer of 15\,nm Au film. Both Au film and magnetic tip were grounded to eliminate any electrostatic interaction between them. A back-gate voltage $V_g$ was applied to the bottom electrode to tune the charge carrier density. The Hall resistance $\rho_{yx}$ and longitudinal resistance $\rho_{xx}$ were obtained by measuring $V_{yx}$ and $V_{xx}$. (b) 5\,K field-dependent MFM images and $\textit{in-situ}$ transport data at $V_g^0 \simeq 10$\,V. The MFM images show ferromagnetic domain behavior during the magnetization reversal from 0.15\,T to 0.35\,T, consistent with transport data.     
\label{fig1}}
\end{figure*}

Indeed, enhanced QAH temperature was observed in Cr and V co-doped BST thin films\cite{ou17}. At optimal Cr/V ratio, full quantization was achieved at 300\,mK, an order of magnitude higher than the end members with single dopants\,\cite{ou17}.  The Hall hysteresis loop is more square-like, suggesting a sharper magnetization reversal, \textit{i.e.}, less magnetic inhomogeneity. Furthermore, the temperature dependence of anomalous Hall resistance is more mean-field-like.  These observations indicate improved ferromagnetism in Cr/V co-doped TI thin films. However, the direct microscopic evidence of long-range ferromagnetic ordering is still lacking. In this letter, we reported a systematic study of Cr/V co-doped BST thin films using the magnetic force microscopy (MFM). Our MFM results reveal a clear ferromagnetic domain behavior of the magnetization reversal process in the optimally doped BST thin films, confirming the long-range ferromagnetic ordering in this QAH system.  Furthermore, the ferromagnetism of co-doped thin films is robust against significant change of bulk charge carrier density, though exchange interaction is enhanced by hole doping.  This indicates a significant contribution from the Ruderman-Kittel-Kasuya-Yosida (RKKY) exchange coupling\,\cite{Ruderman54,kou13}. The direct evidence of long-range ferromagnetic order eases the concern of the fragility of QAHE due to magnetic inhomogeneity, alleviating the need of ultra-low temperature to achieve full quantization.  Our results will encourage further exploration of QAHE  in magnetically doped topological materials for dissipationless conduction at elevated temperature.   

\begin{figure*}[htbp]
\includegraphics[width=\textwidth ]{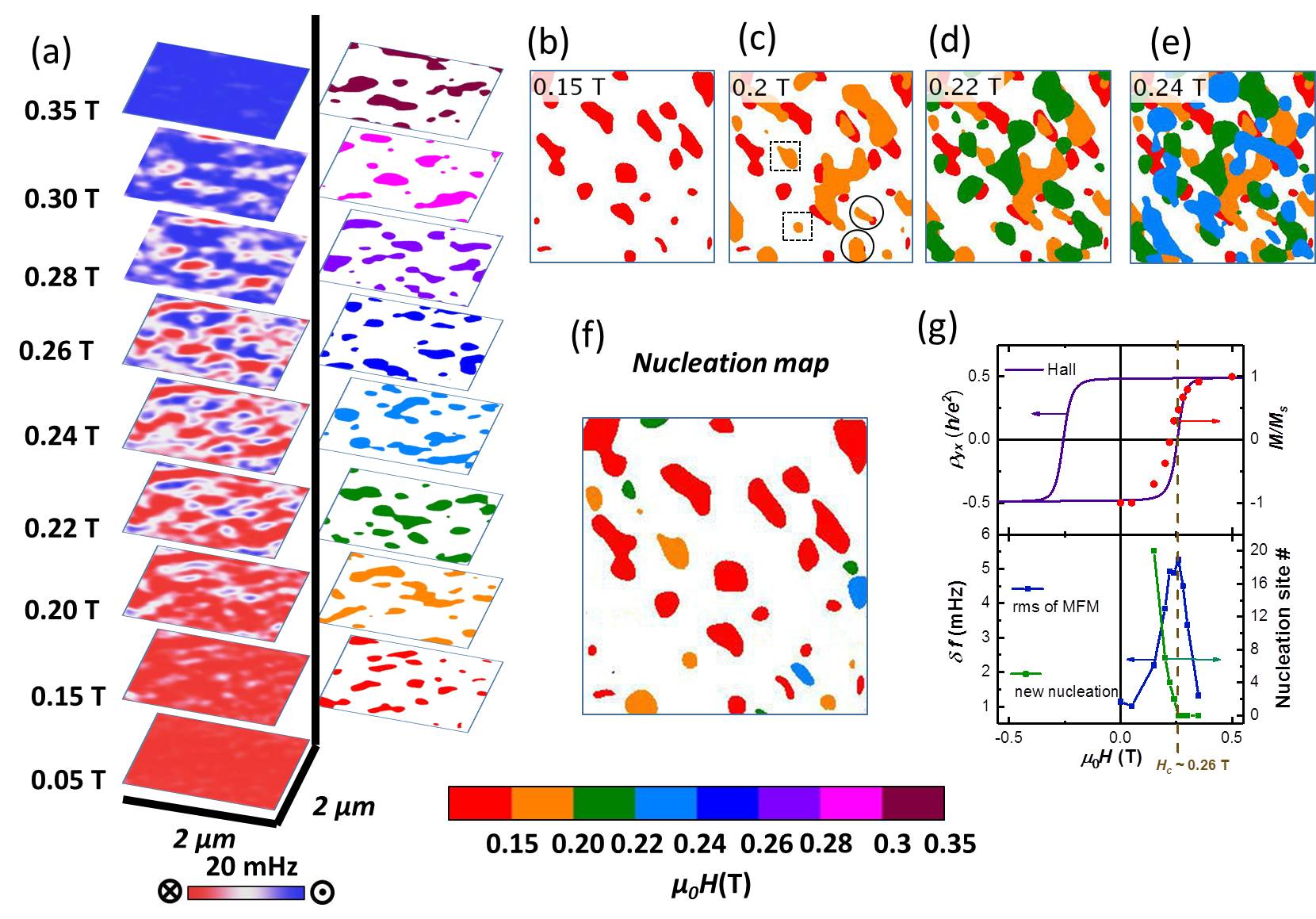}
\caption{\textbf{The magnetization reversal process at 5\,K at neutral point $V_g^0$.} $|$ (a) Left column: a stack of MFM images from 0.05\,T to 0.35\,T to illustrate the domain evolution; Right column: the differential images by taking the difference between adjacent MFM images. Different colors represent local magnetization reversed at different magnetic fields. (b)-(e) Images of the reversed areas at various fields by combining the differential images, which illustrate domain nucleation and domain wall propagation. Dashed squares label the nucleation sites and solid circles label the domain wall propagation. (f) Nucleation map shows the distribution of nucleation sites appearing at various fields. (g) $H$-dependence of normalized magnetization $M/M_s$, estimated from domain population, is quantitatively consistent with the anomalous Hall loop. The domain contrast ($\delta f_{rms}$) peaks at $H_c$, while number of nucleation sites reduces to zero at $H_c$. 
\label{fig2}}
\end{figure*}

Figure\,\ref{fig1}(a) shows the schematic picture of the Hall bar device of the magnetic TI thin films fabricated for MFM and $\textit{in-situ}$ transport measurements.  Three (Cr$_y$V$_{1-y}$)$_{0.19}$(Bi$_x$Sb$_{1-x}$)$_{1.81}$Te$_3$ films ($y = 0, 0.16, 1$, respectively and $x\sim0.4$) are fabricated into the Hall bar devices for MFM and $\textit{in-situ}$ transport measurements. The $y = 0.16$ film is the optimized co-doped sample, which exhibits the best ferromagnetic behavior and the highest QAH temperature among the co-doped BST films\,\cite{ou17}.  The temperature dependence of the Hall resistance shows a mean-field-like behavior with the Curie temperature $T_\mathrm{C}\approx 28$\,K, indicating a robust ferromagnetism. The longitudinal resistance starts to decrease right below $T_\mathrm{C}$, indicating that the sample enters the QAH regime as soon as the long-range ferromagnetic order forms (Supplementary Fig.\,S1(a)). The end member ($y = 0$ or 1), however, enters the QAH regime at a much lower temperature. A gate voltage was applied to the back of STO substrate to tune the Fermi level. At 1.5\,K, the Hall resistance reaches $0.95\,h/e^2$ at $V_g^0$ (Supplementary Fig.\,S1). Such a quantization level was only be achieved below $50$\,mK in singly Cr- or V-doped thin films\cite{chang13,checkelsky14, Kou14prl,chang15}. All MFM images presented here were taken at 5\,K, the base temperature of our MFM system. The Hall traces at both 1.5\,K and 5\,K  show similar square-like hysteresis loop, suggesting no qualitative difference between these two temperatures. Furthermore, a larger coercive field ($H_c$) with sharper reversal was observed at 1.5\,K, suggesting better ferromagnetism at the lower temperature. Therefore, the observed ferromagnetic behavior is expected to persist at the lower temperature where full quantization was observed on the same sample\,\cite{ou17}. 

Fig.\,\ref{fig1}(b) shows the MFM images and $\textit{in-situ}$ transport data ($\rho_{xx}$ and $\rho_{yx}$) of  optimally doped film at $V_g^0\simeq$ 10\,V at various magnetic fields. The $\rho_{yx}(H)$ loop shows a saturation $\sim 0.5\,h/e^2$ with a coercive field $H_c \sim 0.26$\,T. The magnetization reversal process from negatively (red) to positively  (blue) magnetized state is illustrated in the MFM images at the bottom of Fig.\,\ref{fig1}(b). The negatively saturated state has very weak magnetic contrast with a small positive field ($+0.05$\,T), indicating a single domain state persist at a small reversed field. The single domain state is static and stable during the course of MFM measurements.  The observed stable single domain state is in sharp contrast to the superparamagnetic behavior previously reported in Cr-doped BST films, where significant magnetic relaxation was already observed at zero magnetic field\,\cite{Lachmane15}. At 0.15\,T, up domains start to nucleate, represented by light blue regions. As the field increases further, up domains expand and down domains shrink. At coercive field $H_c$ where $\rho_{yx}=0$ and $\rho_{xx}$ peaks at $\sim0.8\,h/e^2$,  equally populated up and down domains were observed, confirming the zero magnetization state ($M=0$). For $H\ge0.35$\,T, no red regions is visible in MFM image, indicating the system is in a saturated (single domain) state. The MFM observation of ferromagnetic domain behavior is in excellent agreement with the \textit{in-situ} transport data, suggesting local observation is representative of the global (bulk) properties.

\begin{figure*}[htbp]
\includegraphics[width=\textwidth]{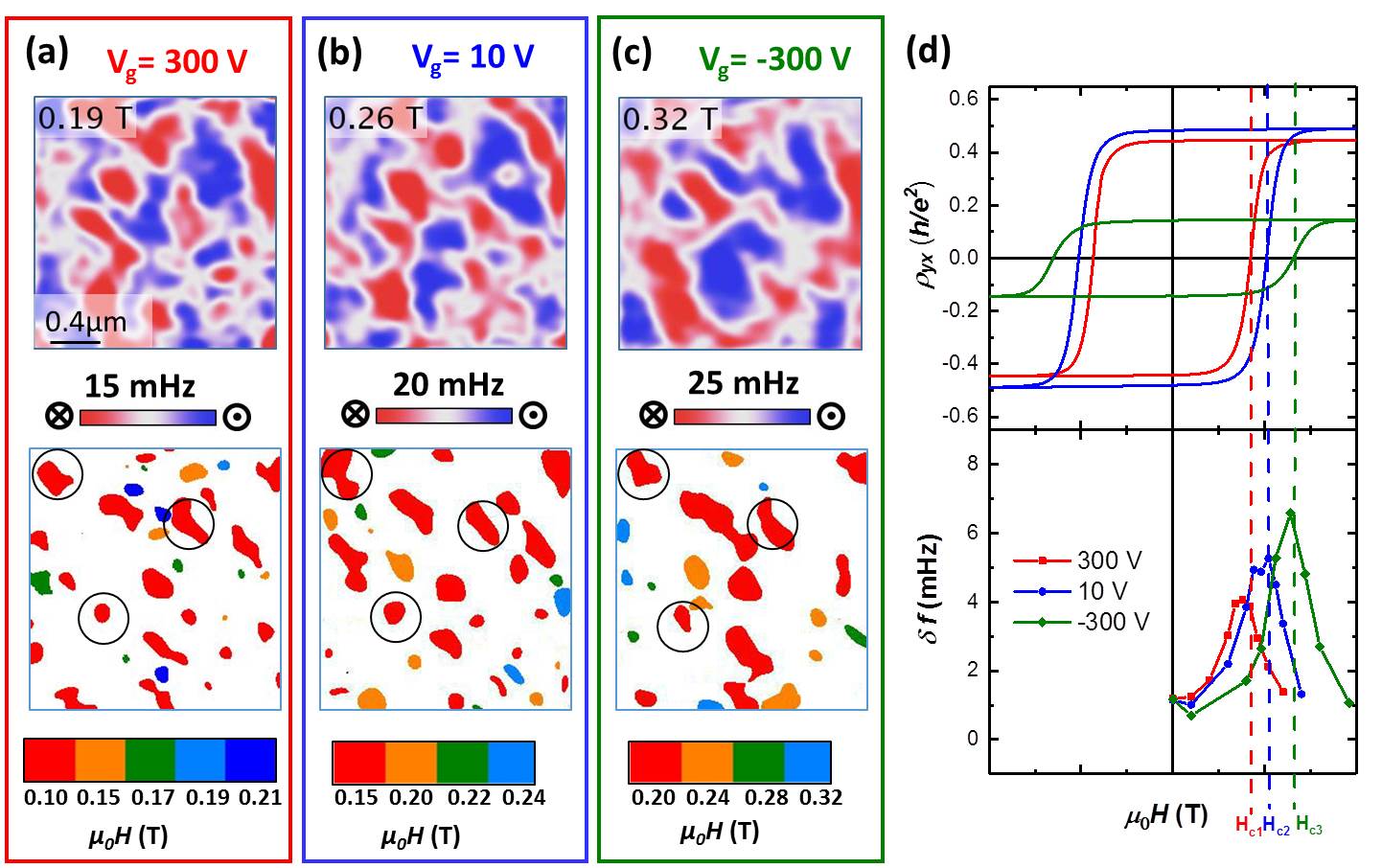}
\caption{\textbf{Gate dependence of ferromagnetic behavior.} $|$ (a)-(c) MFM images around coercive field and nucleation-site maps, respectively at $V_g$ = 300\,V, 10\,V and $-$300\,V. Larger domain size and stronger domain contrast were observed in the $V_g = -300$\,V (hole-doping). Black circles label some of the common nucleation sites at three different $V_g$ values. (d) The Hall resistance (top panel), MFM domain contrasts ($\delta f_{rms}$) (bottom panel) as a function of magnetic fields at three different $V_g$ values. The $H_c$ deduced from two panels are consistent with each other, as $H_{c1}$ $\approx$ 0.21\,T ($V_g$ = 300\,V), $H_{c2}$ $\approx$ 0.26\,T ($V_g^0$ = 10\,V), $H_{c3}$ $\approx 0.33$\,T ($V_g = -300$\,V).    
\label{fig3}}
\end{figure*}

The ferromagnetic domain behavior can be further illustrated with the differential MFM images, the difference between MFM images of adjacent fields as shown in Fig.\,\ref{fig2}(a).  The locations where the changes of MFM signal are above the noise level ($\sim2$\,mHz) are defined as the newly reversed regions.  They are marked by different colors at each field value, which are shown in the right column of Fig.\,\ref{fig2}(a). Here the unchanged areas are marked with white color. In Figs.\,\ref{fig2}(b)-(e), these differential images were stacked together to show the spatial correlations of magnetization reversal events to differentiate domain nucleation from domain expansion. (See supplementary information Fig.\,S4 for complete data set) For example, as shown in Fig.\,\ref{fig2}(c), some of the newly reversed regions (yellow) at 0.2\,T have no correlations with previously reversed regions (red). These regions, labeled by dashed squares, are new nucleation sites. The other yellow regions, labeled by solid circles, have clear overlap with the red regions, demonstrating domain growth due to domain wall propagation. All nucleation sites during the reversal process are summarized in Fig.\,\ref{fig2}(f).  As summarized in Fig.\,\ref{fig2}(g), the number of nucleation sites decreases with increasing $H$, and drops to zero for $H>H_c$, indicating a crossover from domain nucleation to domain wall propagation.   

In addition to direct visualization of ferromagnetic domain behavior, MFM data can also be used to extract the hysteresis loop of normalized magnetization ($M/M_s$), which is estimated from the populations of up and down domains\,\cite{wang16}. As shown in Fig.\,\ref{fig2}(k), the $M/M_s(H)$ curve quantitatively agrees  with $\rho_{yx}(H)$ loop, confirming conventional scaling behavior of anomalous Hall effect: $\rho_{yx}\propto M$. The agreement between local ($M$) and global ($\rho_{yx}$) measurements demonstrate that our MFM results are representative of the bulk magnetic properties.  Consistently, the domain contrast, estimated with the root-mean-square (RMS) value of MFM signal ($\delta f_{rms}$), peaks at $H_c$ when up and down domains are equally populated. The observed domain behavior provides unambiguous evidence of long-range ferromagnetic order in the optimally Cr/V co-doped BST thin films. In contrast, MFM measurements on singly doped BST films do not reveal clear ferromagnetic behavior. (See supplementary information for MFM results of end members) Therefore, our MFM data provide direct evidence that long-range ferromagnetic order is essential for the enhancement of QAH temperature\,\cite{ou17}.

The long-range ferromagnetic order is one of key ingredients of QAHE.  Yu \textit{et al.} proposes van Vleck mechanism in magnetically doped TIs, which gains some experimental support\,\cite{yu10, li2015}.  However, other studies indicate that RKKY type exchange mechanism plays a significant role\cite{li2012,chang15b}.  To shed light on the exchange mechanisms,  we investigate the bulk carriers dependence of the ferromagnetism by applying gate voltage ($V_g$).   Similar to the neutral point $V_g^0$ case, both electron-doped (300\,V) and hole-doped  ($-300$\,V) states  show typical ferromagnetic domain behavior, confirming that long range ferromagnetic order is robust against tuning of Fermi level near neutral point (see Supplementary Information Fig.\,S5 and Fig.\,S6). Fig.\ref{fig3}(a)-(c) show the MFM images at $H_c$ and the nucleation maps of the three gate voltages.  Comparing the three multi-domains states, the hole-doping results in larger domain size, fewer nucleation sites, and stronger domain contrast, while electron doping results in opposite trend. Consistently, $H_c$ is enhanced (suppressed) by hole (electron) doping, as shown in the $\rho_{yx}(H)$ loops in Fig.\,\ref{fig3}(d).  On the other hand, both hole and electron doping away from neutral point suppresses anomalous Hall effect.   Note that domain contrast of multi-domain state is proportional to saturated magnetization\,\cite{wang17}. The enhanced domain contrast at $H_c$ indicate an increase of saturated magnetization with hole doping.   Comparing to the local magnetic moment density, the gate-induced charge carrier density is negligible. (See supplementary information for estimation of induced charge carrier density)   So the saturated magnetization at zero temperature $M_s(0)$ is unlikely to change within our gating capability.  Therefore, the enhancement of the magnetization at 5\,K indicates a decrease of reduced temperature $T/T_\mathrm{C}$, \textit{i.e.}, \textit{i.e.}, higher $T_\mathrm{C}$ due to an enhancement of exchange coupling in the hole-doped films.  This is consistent with higher $H_c$ and fewer nucleation sites in the hole-doped state. In principle, the Dirac surface state carriers can mediate an  RKKY-type exchange interaction\,\cite{Sessi14}. This mechanism is likely irrelevant in our samples because the ferromagnetism gets weaker on the electron-doping side.    Interestingly some nucleation sites (labeled by black circles) are independent of gate voltages, so they  are  likely caused by neutral defects or imperfections that are insensitive to charge carriers.  On the other hand, there are nucleation sites that do depend on gate voltages, which indicates that they might be related to charged defects. 

\begin{figure}[htbp]
\includegraphics[width=\columnwidth ]{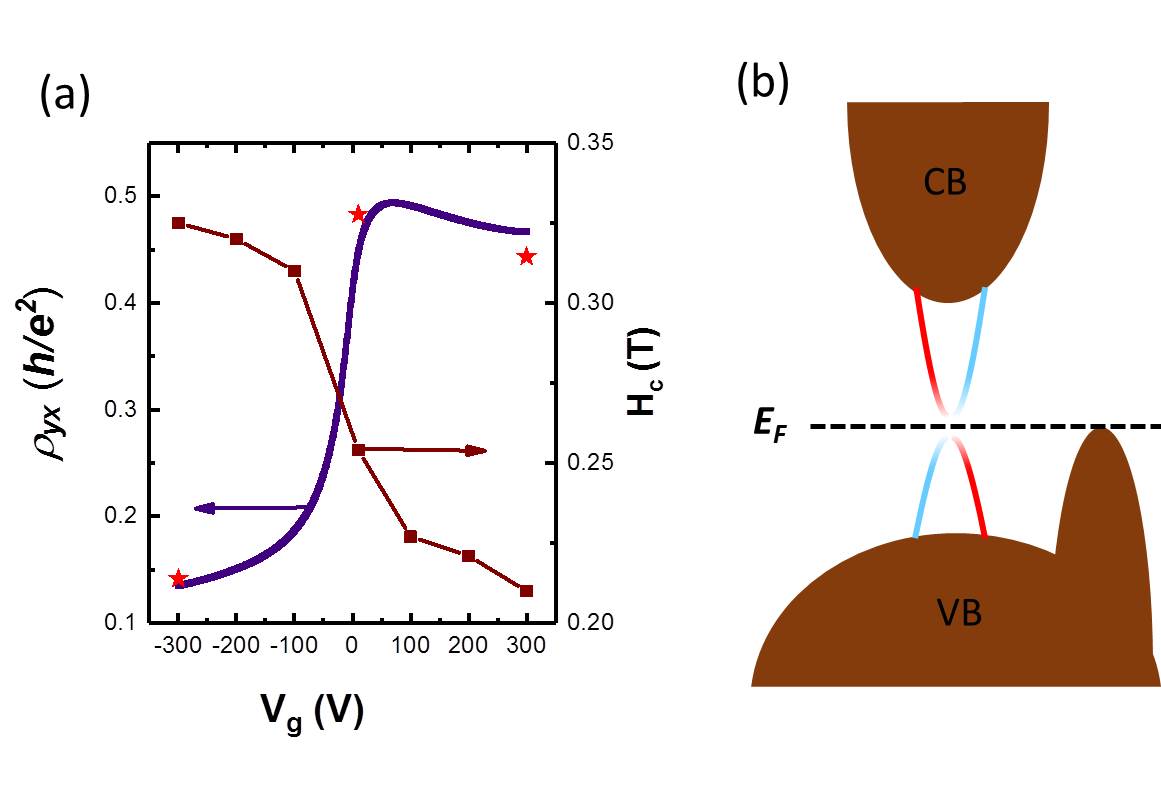}
\caption{\textbf{Gate dependence of the transport and magnetic properties.} $|$ (a) $\rho_{yx}$ and $H_C$ vs. $V_g$. Red stars are the zero-field $\rho_{yx}$ from hysteresis loops at $-$300\,V, 10\,V and 300\,V. (b) The schematic picture of the band structure of the Cr/V co-doped BST film. The Dirac point of the surface state is close to VBM.  
\label{fig4}}
\end{figure}

The $V_g$-dependence of $\rho_{yx}$ and $H_c$ are summarized in Fig.\,\ref{fig4}(a). The $\rho_{yx}(V_g)$ curve was measured at zero magnetic field with slowly ramping of $V_g$ from $-300$\,V to $+300$\,V after the film was saturated at high field. The result agrees well with the saturation state resistance from hysteresis loop measurements (red stars), indicating the film stays in the single-domain state.  The $\rho_{yx}(V_g)$ shows a peak at $V_g^0 \approx$ 10\,V, the charge neutral point. In contrast to the non-monotonic gate dependence of $\rho_{yx}$, the $H_c(V_g)$ shows a monotonic behavior with significant enhancement on hole doping side, consistent with RKKY mechanism. This indicates that the VBM (or the mobility edge) is very close to the neutral (Dirac) point as shown in Fig.\,\ref{fig4}(b)\,\cite{chang15}. Therefore, hole-doping will introduce substantial amount of bulk charge carriers, which will favor RKKY type exchange mechanism but significantly enhance dissipation.   On the other hand, electron-doping push the Fermi level up above the VBM, but presumably not enough to reach the conduction band minimum (CBM). Therefore, electron-doping likely only introduce small amounts of surface state electrons, which slightly enhance dissipation.  Previous ARPES results of V-doped BST film suggest that the Dirac point is below the VBM\,\cite{li16}. Thus doping a small amount of Cr into V-doped BST film might slightly lower the VBM. Alternatively, Cr/V co-doping effectively enhance scattering of bulk carriers, promoting localization without reducing the mass gap.

In conclusion, we present a systematic MFM with \textit{in-siut} transport study of the ferromagnetic domain behavior of Cr/V co-doped BST thin films, which can achieve full quantization of Hall resistance at 300\,mT. MFM results reveal clear domain nucleation and domain wall propagation during the magnetization reversal. Our results provide unambiguous evidence of the long-range ferromagnetic order in magnetic TI thin films. The gate voltage dependence of the ferromagnetism suggests that significant RKKY contribution due to bulk carriers.  The observed robust ferromagnetism resolves one of the major concern in magnetically doped topological insulators, opens a door to explore high-temperature dissipationless conduction with magnetic topological materials.

\section*{Methods}

\subsection{Sample preparation}

Epitaxial thin film Cr/V co-doped BST films capped with 2\,nm Al was grown on heat treated SrTiO$_3$(111) substrate by co-evaporation in a molecular beam epitaxy (MBE) system. The nominal thickness of the film is 5 quintuple layers (QL). The film was scratched by hand into a Hall bar shape connected with a large square-like area for MFM measurements. A layer of $\sim$ 15\,nm Au film was deposited on the square area to eliminated electrostatic interaction between the sample and magnetic tip. 

\subsection{MFM measurement}

The MFM experiments were carried out in a homemade cryogenic atomic force microscope (AFM) using commercial piezoresistive cantilevers (spring constant $k \approx 3$\,N/m, resonant frequency $f_{0} \approx 42\,{\rm kHz}$).  The homemade AFM is interfaced with a Nanonis SPM Controller (SPECS) and a commercial phase-lock loop (SPECS)\cite{wang15, wang16}. MFM tips were prepared by depositing nominally 100 nm Co film onto bare tips using e-beam evaporation. MFM images were taken in a constant mode with the scanning plane $\sim\,40\,{\rm nm}$ above the sample surface. To avoid the relaxation effect (domain wall creeping) near $H_c$ and to minimize the stray field effect of the MFM tip, all MFM images were taken at low magnetic field $\sim$ 0.05\,T after the magnetic field was ramped to the desired values. The MFM signal, the change of cantilever  resonant frequency, is proportional to out-of-plane stray field gradient\cite{rugar90}. Electrostatic interaction was minimized by nulling the tip-surface contact potential difference. Blue (red) regions in MFM images represent up (down) ferromagnetic domains, where magnetizations are parallel(anti-parallel) with the positive external field. 

\subsection{\textit{in-situ} transport measurement}
The magnetic TI films were fabricated into Hall bar devices. The Hall resistance and longitudinal resistance were measured by standard lock-in technique with ac current of 5\,$\mu$A modulated at 314 Hz. 

\section*{Author contribution}
WWu, KH and YW conceived the project.  WWu and WWa designed the MFM experiment.  WWa performed MFM with \textit{in-situ} transport measurements, and analyzed the data.  YO synthesized the MBE films under the supervision of KH and QX.  CL and YW carried out transport characterization of the films.  WWu and WWa wrote the manuscript with inputs from all authors. 

\begin{acknowledgments}
We thank C. Chang for helpful discussion.  This work at Rutgers is supported by the Office of Basic Energy Sciences, Division of Materials Sciences and Engineering, U.S. Department of Energy under Award number DE-SC0008147 and DE-SC0018153.  The work at Tsinghua University is supported by National Natural Science Foundation of China and the Ministry of Science and Technology of China.
\end{acknowledgments}


\begin{thebibliography}{10}
\expandafter\ifx\csname url\endcsname\relax
  \def\url#1{\texttt{#1}}\fi
\expandafter\ifx\csname urlprefix\endcsname\relax\def\urlprefix{URL }\fi
\providecommand{\bibinfo}[2]{#2}
\providecommand{\eprint}[2][]{\url{#2}}

\bibitem{haldane88}
\bibinfo{author}{Haldane, F. D.~M.}
\newblock \bibinfo{title}{Model for a quantum {Hall} effect without landau
  levels: condensed-matter realization of the ``parity anomaly''}.
\newblock \emph{\bibinfo{journal}{Phys. Rev. Lett.}}
  \textbf{\bibinfo{volume}{61}}, \bibinfo{pages}{2015--2018}
  (\bibinfo{year}{1988}).

\bibitem{onoda03}
\bibinfo{author}{Onoda, M.} \& \bibinfo{author}{Nagaosa, N.}
\newblock \bibinfo{title}{Quantized anomalous {Hall} effect in two-dimensional
  ferromagnets: {Quantum} {Hall} effect in metals}.
\newblock \emph{\bibinfo{journal}{Phys. Rev. Lett.}}
  \textbf{\bibinfo{volume}{90}}, \bibinfo{pages}{206601}
  (\bibinfo{year}{2003}).

\bibitem{liu08}
\bibinfo{author}{Liu, C.~X.}, \bibinfo{author}{Qi, X.~L.},
  \bibinfo{author}{Dai, X.}, \bibinfo{author}{Fang, Z.} \&
  \bibinfo{author}{Zhang, S.~C.}
\newblock \bibinfo{title}{Quantum anomalous {Hall} effect in
  {Hg$_{1-y}$Mn$_{1-y}$Te} quantum wells}.
\newblock \emph{\bibinfo{journal}{Phys. Rev. Lett.}}
  \textbf{\bibinfo{volume}{101}}, \bibinfo{pages}{146802}
  (\bibinfo{year}{2008}).

\bibitem{qi08}
\bibinfo{author}{Qi, X.~L.}, \bibinfo{author}{Hughes, T.~L.} \&
  \bibinfo{author}{Zhang, S.~C.}
\newblock \bibinfo{title}{Topological field theory of time-reversal invariant
  insulators}.
\newblock \emph{\bibinfo{journal}{Phys. Rev. B}} \textbf{\bibinfo{volume}{78}},
  \bibinfo{pages}{195424} (\bibinfo{year}{2008}).

\bibitem{yu10}
\bibinfo{author}{Yu, R.} \emph{et~al.}
\newblock \bibinfo{title}{Quantized anomalous {Hall} effect in magnetic
  topological insulators}.
\newblock \emph{\bibinfo{journal}{Science}} \textbf{\bibinfo{volume}{329}},
  \bibinfo{pages}{61--64} (\bibinfo{year}{2010}).

\bibitem{qiao10}
\bibinfo{author}{Qiao, Z.~H.} \emph{et~al.}
\newblock \bibinfo{title}{Quantum anomalous {Hall} effect in graphene from
  {Rashba}and exchange effects}.
\newblock \emph{\bibinfo{journal}{Phys. Rev. B}} \textbf{\bibinfo{volume}{82}},
  \bibinfo{pages}{161414} (\bibinfo{year}{2010}).

\bibitem{nomura11}
\bibinfo{author}{Nomura, K.} \& \bibinfo{author}{Nagaosa, N.}
\newblock \bibinfo{title}{Surface-quantized anomalous {Hall} current and the
  magnetoelectric effect in magnetically disordered topological insulators}.
\newblock \emph{\bibinfo{journal}{Phys. Rev. Lett.}}
  \textbf{\bibinfo{volume}{106}}, \bibinfo{pages}{166802}
  (\bibinfo{year}{2011}).

\bibitem{zhang12}
\bibinfo{author}{Zhang, H.}, \bibinfo{author}{Lazo, C.},
  \bibinfo{author}{Bluegel, S.}, \bibinfo{author}{Heinze, S.} \&
  \bibinfo{author}{Mokrousov, Y.}
\newblock \bibinfo{title}{Electrically tunable quantum anomalous hall effect in
  graphene decorated by 5d transition-metal adatoms}.
\newblock \emph{\bibinfo{journal}{Phys. Rev. Lett.}}
  \textbf{\bibinfo{volume}{108}}, \bibinfo{pages}{056802}
  (\bibinfo{year}{2012}).

\bibitem{ezawa12}
\bibinfo{author}{Ezawa, M.}
\newblock \bibinfo{title}{Valley-polarized metals and quantum anomalous
  {Hall}effect in silicene}.
\newblock \emph{\bibinfo{journal}{Phys. Rev. Lett.}}
  \textbf{\bibinfo{volume}{109}}, \bibinfo{pages}{055502}
  (\bibinfo{year}{2012}).

\bibitem{chang13}
\bibinfo{author}{Chang, C.-Z.} \emph{et~al.}
\newblock \bibinfo{title}{Experimental observation of the quantum anomalous
  {Hall} effect in a magnetic topological insulator}.
\newblock \emph{\bibinfo{journal}{Science}} \textbf{\bibinfo{volume}{340}},
  \bibinfo{pages}{167--170} (\bibinfo{year}{2013}).

\bibitem{checkelsky14}
\bibinfo{author}{Checkelsky, J.~G.} \emph{et~al.}
\newblock \bibinfo{title}{Trajectory of the anomalous {Hall} effect towards the
  quantized state in a ferromagnetic topological insulator}.
\newblock \emph{\bibinfo{journal}{Nat. Phys.}} \textbf{\bibinfo{volume}{10}},
  \bibinfo{pages}{731} (\bibinfo{year}{2014}).

\bibitem{Kou14prl}
\bibinfo{author}{Kou, X.} \emph{et~al.}
\newblock \bibinfo{title}{Scale-invariant quantum anomalous {Hall} effect in
  magnetic topological insulators beyond the two-dimensional limit}.
\newblock \emph{\bibinfo{journal}{Phys. Rev. Lett.}}
  \textbf{\bibinfo{volume}{113}}, \bibinfo{pages}{137201}
  (\bibinfo{year}{2014}).

\bibitem{kou15}
\bibinfo{author}{Kou, X.} \emph{et~al.}
\newblock \bibinfo{title}{Metal-to-insulator switching in quantum anomalous
  {Hall}states}.
\newblock \emph{\bibinfo{journal}{Nat. Comm.}} \textbf{\bibinfo{volume}{6}},
  \bibinfo{pages}{8474} (\bibinfo{year}{2015}).

\bibitem{feng15}
\bibinfo{author}{Feng, Y.} \emph{et~al.}
\newblock \bibinfo{title}{Observation of the zero {Hall} plateau in a quantum
  anomalous {Hall} insulator}.
\newblock \emph{\bibinfo{journal}{Phys. Rev. Lett.}}
  \textbf{\bibinfo{volume}{115}}, \bibinfo{pages}{126801}
  (\bibinfo{year}{2015}).

\bibitem{mogi15}
\bibinfo{author}{Mogi, M.} \emph{et~al.}
\newblock \bibinfo{title}{Magnetic modulation doping in topological insulators
  toward higher-temperature quantum anomalous {Hall} effect}.
\newblock \emph{\bibinfo{journal}{Appl. Phys. Lett.}}
  \textbf{\bibinfo{volume}{107}}, \bibinfo{pages}{182401}
  (\bibinfo{year}{2015}).

\bibitem{ou17}
\bibinfo{author}{Ou, Y.} \emph{et~al.}
\newblock \bibinfo{title}{Enhancing the quantum anomalous hall effect by
  magnetic co-doping in a topological insulator} (\bibinfo{year}{2017}).
\newblock \bibinfo{note}{Submitted}.

\bibitem{Lachmane15}
\bibinfo{author}{Lachman, E.~O.} \emph{et~al.}
\newblock \bibinfo{title}{Visualization of superparamagnetic dynamics in
  magnetic topological insulators}.
\newblock \emph{\bibinfo{journal}{Sci. Adv.}} \textbf{\bibinfo{volume}{1}},
  \bibinfo{pages}{e1500740} (\bibinfo{year}{2015}).

\bibitem{grauer15}
\bibinfo{author}{Grauer, S.} \emph{et~al.}
\newblock \bibinfo{title}{Coincidence of superparamagnetism and perfect
  quantization in the quantum anomalous {Hall} state}.
\newblock \emph{\bibinfo{journal}{Phys. Rev. B}} \textbf{\bibinfo{volume}{92}},
  \bibinfo{pages}{201304} (\bibinfo{year}{2015}).

\bibitem{lee15}
\bibinfo{author}{Lee, I.} \emph{et~al.}
\newblock \bibinfo{title}{Imaging {Dirac-mass} disorder from magnetic dopant
  atoms in the ferromagnetic topological insulator
  {Cr$_x$(Bi$_{0.1}$Sb$_{0.9}$)$_{2-x}$Te$_3$}}.
\newblock \emph{\bibinfo{journal}{Proc. Natl. Acad. Sci. U. S. A.}}
  \textbf{\bibinfo{volume}{112}}, \bibinfo{pages}{1316--1321}
  (\bibinfo{year}{2015}).

\bibitem{bednorz86}
\bibinfo{author}{Bednorz, J.~G.} \& \bibinfo{author}{Muller, K.~A.}
\newblock \bibinfo{title}{Possible high {T$_{c}$} superconductivity in the
  {Ba-La-Cu-O} system}.
\newblock \emph{\bibinfo{journal}{Z. Fur Physik B-condensed Matter}}
  \textbf{\bibinfo{volume}{64}}, \bibinfo{pages}{189--193}
  (\bibinfo{year}{1986}).

\bibitem{wu87}
\bibinfo{author}{Wu, M.~K.} \emph{et~al.}
\newblock \bibinfo{title}{Superconductivity at 93 {K} in a new mixed-phase
  {Y-Ba-Cu-O} compound system at ambient pressure}.
\newblock \emph{\bibinfo{journal}{Phys. Rev. Lett.}}
  \textbf{\bibinfo{volume}{58}}, \bibinfo{pages}{908--910}
  (\bibinfo{year}{1987}).

\bibitem{maeda88}
\bibinfo{author}{Maeda, H.}, \bibinfo{author}{Tanaka, Y.},
  \bibinfo{author}{Fukutomi, M.} \& \bibinfo{author}{Asano, T.}
\newblock \bibinfo{title}{A new high-{T$_{c}$} oxide superconductor without a
  rare earth element}.
\newblock \emph{\bibinfo{journal}{Jpn. J. Appl. Phys., Part 2}}
  \textbf{\bibinfo{volume}{27}}, \bibinfo{pages}{L209--L210}
  (\bibinfo{year}{1988}).

\bibitem{schilling93}
\bibinfo{author}{Schilling, A.}, \bibinfo{author}{Cantoni, M.},
  \bibinfo{author}{Guo, J.~D.} \& \bibinfo{author}{Ott, H.~R.}
\newblock \bibinfo{title}{Superconductivity above 130 {K} in the
  {Hg-Ba-Ca-Cu-O} system}.
\newblock \emph{\bibinfo{journal}{Nature}} \textbf{\bibinfo{volume}{363}},
  \bibinfo{pages}{56--58} (\bibinfo{year}{1993}).

\bibitem{chang15}
\bibinfo{author}{Chang, C.~Z.} \emph{et~al.}
\newblock \bibinfo{title}{High-precision realization of robust quantum
  anomalous {Hall} state in a hard ferromagnetic topological insulator}.
\newblock \emph{\bibinfo{journal}{Nat. Mater.}} \textbf{\bibinfo{volume}{14}},
  \bibinfo{pages}{473--477} (\bibinfo{year}{2015}).

\bibitem{Grauer2017}
\bibinfo{author}{Grauer, S.} \emph{et~al.}
\newblock \bibinfo{title}{Scaling of the quantum anomalous {Hall} effect as an
  indicator of axion electrodynamics}.
\newblock \emph{\bibinfo{journal}{Phys. Rev. Lett.}}
  \textbf{\bibinfo{volume}{118}}, \bibinfo{pages}{246801}
  (\bibinfo{year}{2017}).

\bibitem{chang14}
\bibinfo{author}{Chang, C.-Z.} \emph{et~al.}
\newblock \bibinfo{title}{Chemical-potential-dependent gap opening at the dirac
  surface states of {${\mathrm{Bi}}_{2}{\mathrm{Se}}_{3}$} induced by
  aggregated substitutional {Cr} atoms}.
\newblock \emph{\bibinfo{journal}{Phys. Rev. Lett.}}
  \textbf{\bibinfo{volume}{112}}, \bibinfo{pages}{056801}
  (\bibinfo{year}{2014}).

\bibitem{li16}
\bibinfo{author}{Li, W.} \emph{et~al.}
\newblock \bibinfo{title}{Origin of the low critical observing temperature of
  the quantum anomalous {Hall} effect in {V-doped} {(Bi,Sb)$_2$Te$_3$} film}.
\newblock \emph{\bibinfo{journal}{Sci. Rep.}} \textbf{\bibinfo{volume}{6}},
  \bibinfo{pages}{32732} (\bibinfo{year}{2016}).

\bibitem{andriotis13}
\bibinfo{author}{Andriotis, A.~N.} \& \bibinfo{author}{Menon, M.}
\newblock \bibinfo{title}{Defect-induced magnetism: {Codoping} and a
  prescription for enhanced magnetism}.
\newblock \emph{\bibinfo{journal}{Phys. Rev. B}} \textbf{\bibinfo{volume}{87}},
  \bibinfo{pages}{155309} (\bibinfo{year}{2013}).

\bibitem{qi16}
\bibinfo{author}{Qi, S.~F.} \emph{et~al.}
\newblock \bibinfo{title}{High-temperature quantum anomalous {Hall} effect in
  n-p codoped topological insulators}.
\newblock \emph{\bibinfo{journal}{Phys. Rev. Lett.}}
  \textbf{\bibinfo{volume}{117}}, \bibinfo{pages}{056804}
  (\bibinfo{year}{2016}).

\bibitem{Ruderman54}
\bibinfo{author}{Ruderman, M.~A.} \& \bibinfo{author}{Kittel, C.}
\newblock \bibinfo{title}{Indirect exchange coupling of nuclear magnetic
  moments by conduction electrons}.
\newblock \emph{\bibinfo{journal}{Phys. Rev.}} \textbf{\bibinfo{volume}{96}},
  \bibinfo{pages}{99--102} (\bibinfo{year}{1954}).

\bibitem{kou13}
\bibinfo{author}{Kou, X.~F.} \emph{et~al.}
\newblock \bibinfo{title}{Interplay between different magnetisms in {Cr}-doped
  topological insulators}.
\newblock \emph{\bibinfo{journal}{Acs Nano}} \textbf{\bibinfo{volume}{7}},
  \bibinfo{pages}{9205--9212} (\bibinfo{year}{2013}).

\bibitem{wang16}
\bibinfo{author}{Wang, W.}, \bibinfo{author}{Chang, C.-Z.},
  \bibinfo{author}{Moodera, J.~S.} \& \bibinfo{author}{Wu, W.}
\newblock \bibinfo{title}{Visualizing ferromagnetic domain behavior of magnetic
  topological insulator thin films}.
\newblock \emph{\bibinfo{journal}{npj Quantum Mater.}}
  \textbf{\bibinfo{volume}{1}}, \bibinfo{pages}{16023} (\bibinfo{year}{2016}).

\bibitem{li2015}
\bibinfo{author}{Li, M.} \emph{et~al.}
\newblock \bibinfo{title}{Experimental verification of the {Van} {Vleck} nature
  of long-range ferromagnetic order in the vanadium-doped three-dimensional
  topological insulator {Sb$_{2}$Te$_{3}$}}.
\newblock \emph{\bibinfo{journal}{Phys. Rev. Lett.}}
  \textbf{\bibinfo{volume}{114}}, \bibinfo{pages}{146802}
  (\bibinfo{year}{2015}).

\bibitem{li2012}
\bibinfo{author}{Li, H.} \emph{et~al.}
\newblock \bibinfo{title}{Carriers dependence of the magnetic properties in
  magnetic topological insulator {Sb$_{1.95-x}$Bi$_x$Cr$_{0.05}$Te$_3$}}.
\newblock \emph{\bibinfo{journal}{Appl. Phys. Lett.}}
  \textbf{\bibinfo{volume}{101}}, \bibinfo{pages}{072406}
  (\bibinfo{year}{2012}).

\bibitem{chang15b}
\bibinfo{author}{Chang, C.-Z.} \emph{et~al.}
\newblock \bibinfo{title}{Zero-field dissipationless chiral edge transport and
  the nature of dissipation in the quantum anomalous hall state}.
\newblock \emph{\bibinfo{journal}{Phys. Rev. Lett.}}
  \textbf{\bibinfo{volume}{115}}, \bibinfo{pages}{057206}
  (\bibinfo{year}{2015}).

\bibitem{wang17}
\bibinfo{author}{Wang, W.} \emph{et~al.}
\newblock \bibinfo{title}{Visualizing weak ferromagnetic domains in
  multiferroic hexagonal ferrite thin film}.
\newblock \emph{\bibinfo{journal}{Phys. Rev. B}} \textbf{\bibinfo{volume}{95}},
  \bibinfo{pages}{134443} (\bibinfo{year}{2017}).

\bibitem{Sessi14}
\bibinfo{author}{Sessi, P.} \emph{et~al.}
\newblock \bibinfo{title}{{Signatures of Dirac fermion-mediated magnetic
  order}}.
\newblock \emph{\bibinfo{journal}{Nat Commun}} \textbf{\bibinfo{volume}{5}},
  \bibinfo{pages}{5349} (\bibinfo{year}{2014}).

\bibitem{wang15}
\bibinfo{author}{Wang, W.} \emph{et~al.}
\newblock \bibinfo{title}{Visualizing ferromagnetic domains in magnetic
  topological insulators}.
\newblock \emph{\bibinfo{journal}{APL Mater.}} \textbf{\bibinfo{volume}{3}},
  \bibinfo{pages}{083301} (\bibinfo{year}{2015}).

\bibitem{rugar90}
\bibinfo{author}{Rugar, D.} \emph{et~al.}
\newblock \bibinfo{title}{Magnetic force microscopy: {General} principles and
  application to longitudinal recording media}.
\newblock \emph{\bibinfo{journal}{J. Appl. Phys.}}
  \textbf{\bibinfo{volume}{68}}, \bibinfo{pages}{1169--1183}
  (\bibinfo{year}{1990}).

\end{thebibliography}
\end{document}